\documentclass[aps,prc,
superscriptaddress,
preprint,
showpacs]{revtex4-1}
\usepackage{amsmath}
\usepackage{color}
\usepackage{graphicx}
\begin{document}
\title{Investigation of $^{23}$N in a three-body model}
\author{Liuyang Zhang} \email{liuyang\_zhang@hotmail.com}
\affiliation{School of Physics and Key Laboratory of Modern Acoustics,
Institute of Acoustics, Nanjing University, Nanjing 210093, China}
\author{Zhongzhou Ren} \email{zren@nju.edu.cn}  \affiliation{School
  of Physics and Key Laboratory of Modern Acoustics, Institute of
  Acoustics, Nanjing University, Nanjing 210093, China}
\affiliation{Center of Theoretical Nuclear Physics, National
  Laboratory of Heavy-Ion Accelerator, Lanzhou 730000, China}
\author{Mengjiao Lyu} \email{mengjiao\_lyu@hotmail.com}
\affiliation{School of Physics and Key Laboratory of Modern Acoustics,
Institute of Acoustics, Nanjing University, Nanjing 210093, China}
\author{Chen Ji} \email{jichen@triumf.ca}  \affiliation{TRIUMF,
4004 Wesbrook Mall, Vancouver, British Columbia V6T 2A3, Canada}

\date{\today}
 
\begin{abstract}
  The neutron-drip-line nucleus $^{23}$N is investigated in a
  three-body model consisting of a $^{21}$N core and two valence
  neutrons. Using the Faddeev formalism with the realistic
  neutron-neutron potential and the neutron-core potential, we
  calculate the ground-state properties of $^{23}$N including its two-neutron separation
  energy, and obtain good agreement with the experiments. We also find an
  excited $^{23}$N state with a shallow two-neutron separation energy at
  about 0.18 MeV. By
  evaluating the root-mean-square matter radii, the average distances
  between the two valence neutrons, and the average distances from the
  core to the center-of-mass of the valence-neutron pair, we show that
  the excited state of $^{23}$N has a distinct halo structure.  Through calculating
  the correlation density distributions of the $^{23}$N three-body system in configuration space,
  we find that the excited state of $^{23}$N has a triangular shape, which is similar to but much more
  extended than the ground state. This scaling symmetry between the ground and excited states indicate 
  the existence of an Efimov state in $^{23}$N.
 \end{abstract}

\pacs{21.10.Gv, 21.60.-n, 27.30.+t}

\maketitle

\section{Introduction}
The properties of neutron-rich nuclei close to the drip-line are
crucial for understanding fundamental mechanism governing the stability and
formation of nuclei. In recent years, tremendous experimental and
theoretical efforts have been devoted to the investigation of
neutron-rich nuclei
\cite{Mitting1987,Saint1989,Bertsch1991,Sakurai1996,Neff2008}.  The
discoveries of light nuclei with halo structures have attracted great
attention because of their exotic properties
\cite{Hansen1987,Ren1990,Zhukov1993,Tanihata1996,Nakamura2009}.
and their important role in the big bang nucleosysnthesis \cite{Rupak2011}.
For the nitrogen isotopes, many studies have focused on neutron-rich
isotopes with mass number ranging from $A=17$ to $A=22$
\cite{Mueller1990,Ozawa1994,Tilley1998,Elekes2010,Hua2011,Rodr2011}. However, 
$^{23}$N has rarely been investigated since its first
production in 1998 \cite{1Audi2012,Yoneda2003}. As a newly synthesized
drip-line nucleus, many physical properties of $^{23}$N such as the
energy spectrum and the structure configuration have not been observed or predicted 
yet except for the one-neutron separation energy $S_{n}$. $S_{n}$s in $^{22}$N and $^{23}$N are respectively
1.28 MeV and 1.79 MeV, which are much smaller than that in $^{21}$N of 4.59 MeV
\cite{3Audi2012}.  Therefore, in a first-order approximation, one can treat the $^{23}$N
nucleus as a three-body system composed by an inert $^{21}$N core,
which is tightly bound, and two valance neutrons moving at a relatively large distance away from the core.

Various few-body methods have been adopted to calculate different three-body quantum systems, such as the Faddeev formalism
\cite{Thompson2004}, the equivalent two-body method \cite{1Ren1994,
  2Ren1994}, the Green's Function Monte Carlo \cite{Lv2012},
and the THSR (Tohsaki-Horiuchi-Schuck-R\"{o}pke) wave function \cite{Tohsaki2001,Lv2014}.  Among these
different approaches, the Faddeev formalism has a specific advantage
that it can also describe the three-body  mechanism in a heavier system, which is normally
encoded in microscopic calculations, in a computationally simpler way \cite{Thompson2004}.  Various neutron-rich nuclei,
including $^{6,8}$He, $^{11}$Li, $^{12,14}$Be,
$^{17}$B, and $^{22}$C have been investigated with the Faddeev
formalism
\cite{Vaagen1994,Nunes1996,Tostevin1997,Thompson2000,Brida2006,Chu2008,Romero2008,Chen2009,Acharya2013,Ji2014}.
This approach has also been applied recently to proton-rich nuclei, such
as $^{17}$Ne, $^{18}$Ne, and $^{28}$S \cite{Zhukov1995,Yanyun2008}.
In this work, we have applied the Faddeev equations to the ${}^{21}$N+n+n three-body system.  By solving these
equations, we obtain various properties of the $^{23}$N nucleus,
which are essential for understanding the halo structure inside the
nucleus.

In Sec.~\ref{sec:Faddeev equations}, we provide a brief introduction to
the Faddeev equations and the two-body interactions adopted in our calculations.  
In Sec.~\ref{sec:num_res}, we
present the numerical results of the three-body calculation and
analyze the halo structure of the ground state and our newly discovered
excited state in $^{23}$N.  At last, the conclusion is given in Sec.~\ref{sec:con_dis}.

\section{FORMALISM}
\label{sec:Faddeev equations}
\begin{figure}[htbp]
  \centering
  \includegraphics[width=0.45\textwidth]{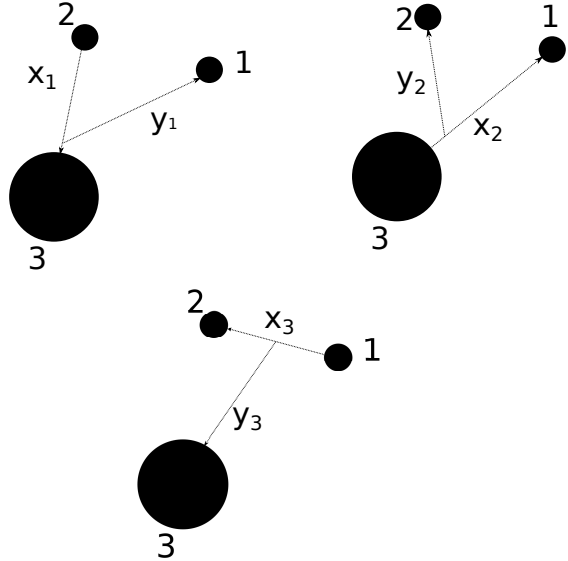}
  \caption{\label{fig:Jacobi coordinates}Three sets of Jacobi
    coordinates in the Faddeev formalism for the core+n+n three-body system.
  }
\end{figure}

In the Faddeev formalism, the full wave function $\Psi^{JM}$ for a three-body system 
can be decomposed into three components with respect to different sets of Jacobi coordinates \cite{Faddeev1960}:
\begin{equation}
\Psi^{JM} =
\Psi_{1}^{JM}(x_{1},y_{1})+\Psi_{2}^{JM}(x_{2},y_{2})+\Psi_{3}^{JM}(x_{3},y_{3}),
\end{equation}
where $\Psi_{i}^{JM}$ is the $i$-th component of the full wave
function with total angular momentum $J$ and its z-component $M$. The i-th set of the 
Jacobi coordinates corresponding to this component is shown in Fig.~\ref{fig:Jacobi coordinates} 
for $i=1,2,3$ \cite{Thompson2004}.
 
The coupled-channel Faddeev equations for the core+n+n system are expressed as
\begin{equation}
\label{eq:three body Faddeev equations} 
\begin{array}{ccc}
 &(T_{1}+V_{1}-E)\Psi_{1}^{JM}=-V_{1}(\Psi_{2}^{JM}+\Psi_{3}^{JM})\\
 &(T_{2}+V_{2}-E)\Psi_{2}^{JM}=-V_{2}(\Psi_{3}^{JM}+\Psi_{1}^{JM})\\ 
 &(T_{3}+V_{3}-E)\Psi_{3}^{JM}=-V_{3}(\Psi_{1}^{JM}+\Psi_{2}^{JM})\\
\end{array} 
\end{equation}
where $T_{i}=T_{xi}+T_{yi}$ is the relative kinetic energy in $i$-th coordinate set.  $V_{i} \equiv V_{jk}$
represents the effective two-body nuclear force between particles $j$ and $k$. 
The indices $(i,j,k) \equiv (1, 2, 3)$ run through cyclic permutation \cite{Thompson2004}.

We then introduce the Jacobi coordinates, which include the hyper radius
$\rho$ and the hyper-angle $\theta_{i}$. They are related to the Jacobi coordinates by \cite{Nunes1996,Thompson2004}
\begin{equation}
\label{eq:hyperspherical coordinates} 
\rho^{2}= x_{i}^{2}+y_{i}^{2} \quad,\quad  \theta_{i}=\text{arctan}(\frac{x_{i}}{y_{i}}).
\end{equation}
One should notice that the hyper-radius is the same in different
sets of Jacobi representation, but the hyper-angles are different. 
To solve the Faddeev equations, we adopt the hyperspherical harmonic method.  Using this approach, 
one can conveniently separate the hyper-angle and hyper-radial dependencies of the wave
function.
The hyperspherical decomposition of the i-th component of the wave function is
defined as  \cite{Nunes1996,Yanyun2008}
\begin{equation}
\label{eq:hyperspherical wave function} 
\Psi_{i}^{JM}=\rho^{-5/2}\sum_{\alpha_{i}}\sum_{K_{i}}\chi_{\alpha_{i}K_{i}}^{i,J}(\rho)\phi_{K_{i}}^{l_{xi}l_{yi}}(\theta_{i})
\left|i:\alpha_{i}\right\rangle,
\end{equation}
where
$\alpha_{i}\equiv{\{l_{xi},l_{yi},L_{i},s_{j},s_{k},S_{xi}\}}J_{i}$
and $\left|i:\alpha_{i}\right\rangle$ represent the quantum numbers in the specific partial-wave channel
occupied by the three-body system. The hyper-angular function
$\phi_{K_{i}}^{l_{xi}l_{yi}}(\theta_{i})$, which are the
eigen-solution of the hyper-angular equation, is expanded in terms of
Jacobi polynomials as
\begin{equation}
\label{eq:hyper-angular wavefunctions} 
\phi_{K_{i}}^{l_{xi}l_{yi}}(\theta_{i})=N_{K_{i}}^{l_{xi}l_{yi}}(\text{sin}\theta_{i})^{L_{xi}}(\text{cos}\theta_{i})^{L_{yi}}P_{n_{i}}^{l_{xi}+1/2,l_{yi}+1/2}(\text{cos}2\theta_{i}).            
\end{equation}
Here $P_{n_{i}}^{l_{xi}+1/2,l_{yi}+1/2}(\text{cos}2\theta_{i})$ denotes the
Jacobi polynomial with $n_{i}=0,1,2,\cdots$. $N_{K_{i}}^{l_{xi}l_{yi}}$ is the normalization
coefficient, and $K_{i}=l_{xi}+l_{yi}+2n_{i}$ indicates the hyper-angular momentum with respect
to the corresponding Jacobi polynomial.

After substituting Eq.~\eqref{eq:hyper-angular wavefunctions} into Eq.~\eqref{eq:hyperspherical wave function}, the
hyper-radial function $\chi_{\alpha_{i}K_{i}}^{i,J}(\rho)$ and the
hyper-angular function $\phi_{K_{i}}^{l_{xi}l_{yi}}(\theta_{i})$ are obtained using the numerical 
program to solve the Faddeev equations~\cite{Thompson2004}.

In the calculations of $^{23}$N as the $^{21}$N+n+n system, the core-neutron
and the neutron-neutron interactions can be determined phenomenologically by fixing the
experimental data.  In this paper, we adopt the well-known GPT
potential neglecting only the spin-spin term for the n-n interaction and keeping the central,
tensor and spin-orbit parts. This potential provides good fits to
the low-energy properties of the low-energy n-n scattering
\cite{Danilin1998,Thompson2000}.  For the core-neutron interaction, we
adopt a Woods-Saxon form \cite{Horiuchi2006,Yanyun2008,Chen2009}
\begin{equation}
\label{eq:Woods-Saxon Formalism} 
V_{\text{n-core}}(r)=\frac{V_{0}}{1+\exp(\frac{r-r_{0}}{a})}
+\frac{V_{\text{so}}}{ra}
  \frac{\exp(\frac{r-r_{0}}{a})}{(1+\exp(\frac{r-r_{0}}{a}))^{2}}\mathbf{L\cdot S},
\end{equation}
where $r_{0}=1.25A^{1/3}$ fm and $a=0.65$ fm \cite{Horiuchi2006,Yanyun2008}. 
$V_{0}$ is the depth of the
Woods-Saxon potential, and $V_{\text{so}}$ represents the strength of the
spin-orbit coupling. $V_{0}$ and $V_{\text{so}}$ can be determined by
fixing the binding energies of the core-neutron two-body
subsystem. We then apply the super symmetric transformation to this
interaction to eliminate the spurious bound states which are forbidden by the Pauli
principle.

\section{Numerical results}
\label{sec:num_res}
We discuss in this section the numerical results for calculating $^{23}$N in a three-body 
 model. We firstly discuss the physical model for the core and  explain our choices 
 of parameterization for the effective n-$^{21}$N
interaction, which are suitable to reproduce the experimental data. Then we solve the Faddeev 
equations and calculate the neutron-separation energies and configuration-space wave functions 
of $^{23}$N. These observables can be utilized to analyze the halo structure of $^{23}$N.

To construct the effective n-$^{21}$N interaction, information about the shell-model occupations 
of the $^{21}$N core and valance neutrons, which can be determined from
experiments, needs to be known. Here we make two assumptions in the shell-model description.
Firstly, we assume that the neutrons inside the $^{21}$N
core occupy the $(0s_{1/2})^{2}$, $(0p_{3/2})^{4}$, $(0p_{1/2})^{2}$
and $(0d_{5/2})^{6}$ orbits, which are forbidden to be occupied by the valence neutrons 
due to the Pauli principle \cite{Thompson2000}. With the lack of the experimental information
about the excited states in $^{21}$N, we
simply neglect these excited states in our calculations as a first-order approximation.
Similar assumptions are also made in the work by other groups, {\it e.g.}, in Refs. \cite{Ogata2013,Yanyun2008,Horiuchi2006}.
With these limitations, we assume that
the valence neutron occupies the $1s_{1/2}$ state in the n-core sub-system of $^{22}$N.

\begin{figure}[htbp]
  \centering
  \includegraphics[width=0.60\textwidth]{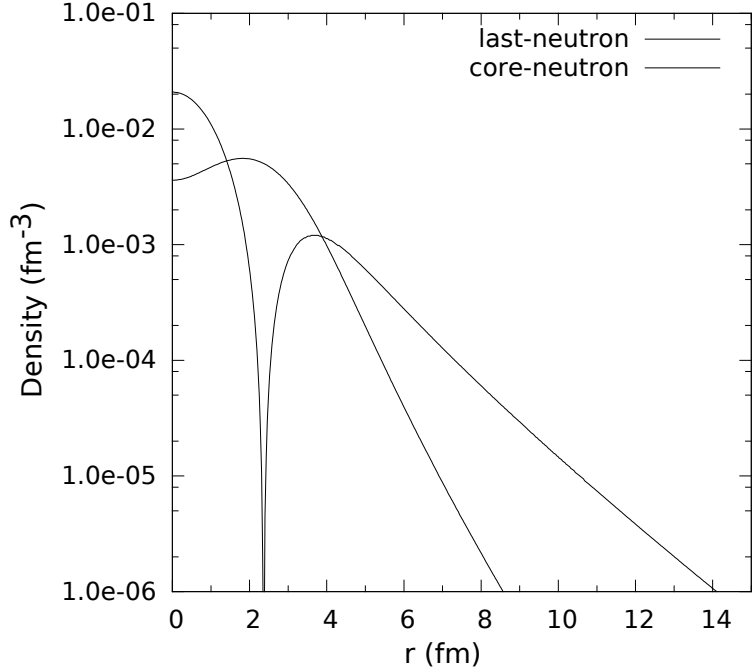}
  \caption{\label{fig:Density distributions}Density distributions of
    neutrons in the subsystem $^{22}$N. The solid curve is the density
    distribution of the core neutrons, and the dashed curve is the
    density distribution of the last neutron. }
\end{figure}

Using the above assumptions and the Wood-Saxon potential form, {\it i.e.}, Eq.~(\ref{eq:Woods-Saxon Formalism}),
we determine the core-n interaction by reproducing two experimental results.  
The first is the
one-neutron separation energy of $^{22}$N, corresponding to the binding of the $1s_{1/2}$ orbit,
{\it i.e.}, $S_{n}=\epsilon(1s_{1/2})=1.28\pm0.21$ MeV. We also take into account the one-neutron 
separation energy of $^{21}$N, which is related to the
binding of the $0d_{5/2}$ orbit, {\it i.e.}, $\epsilon(0d_{5/2})=4.59\pm0.11$ MeV.
The experimental error of $\epsilon(1s_{1/2})$ is considered in our
calculations; while the error bar of $\epsilon(0d_{5/2})$ is neglected due to its relatively small size.
By reproducing the upper bound, lower bound and mean value of $\epsilon(1s_{1/2})$, and also fixing the
mean value of $\epsilon(0d_{5/2})$, we determine three sets of parameters of 
the core-n interactions and list them in Table \ref{table:parameters}. 
The corresponding root-mean-square (r.m.s.) radii of the core, the valence neutron and total matter in the $^{21}$N-n
two-body system are calculated with respect to different parameterizations and shown in Table
\ref{table:properties}. 
The neutron density distributions of $^{22}$N are shown in
Fig.~\ref{fig:Density distributions}, where the probability density of the neutrons inside 
core and of the last neutron occupying the $1s_{1/2}$ state, are shown respectively. We find
that the last neutron has a much more extended matter density distribution
than the neutrons inside the core do. This clearly indicates a halo structure in the $^{21}$N-n system.

\begin{table*}[htbp]
  \begin{center}
    \caption{\label{table:parameters}Parameters of the effective
      $^{21}$N-n interaction.  $\epsilon$ is the single-particle energy of
      the $^{21}$N-n system. The upper bound (1.49 MeV), the mean value (1.28 MeV), 
      and the lower bound (1.07) MeV of $S_{n}$ in
      $^{22}$N are adopted for sets A, B, and C, respectively.}
  \begin{tabular}{l  c  c  c  c  c  c  c  c  c  }
    \hline
    \hline
   &$r_{0}$  &$a$  &$V_{0}$ &$V_{\text{so}}$ &$\epsilon(0s_{1/2})$ &$\epsilon(0p_{3/2})$ &$\epsilon(0p_{1/2})$ &$\epsilon(0d_{5/2})$ &$\epsilon(1s_{1/2})$  \\
   &(fm)     &(fm)  &(MeV) &(MeV.fm$^{2}$) &(MeV) &(MeV) &(MeV) &(MeV) &(MeV)\\
   \hline
   Set A &3.45 &0.60  &-41.45 &-46.45 &-25.382 &-14.939 &-9.026 &-4.590 &-1.490 \\
   Set B &3.45 &0.60  &-40.77 &-50.51 &-24.813 &-14.647 &-8.224 &-4.590 &-1.280 \\
   Set C &3.45 &0.60  &-40.02 &-55.00 &-24.170 &-14.316 &-7.333 &-4.590 &-1.070 \\
    \hline
    \hline
  \end{tabular}
  \end{center}
\end{table*}

\begin{table*}[htbp]
  \begin{center}
    \caption{\label{table:properties}The calculated r.m.s.
      radii of the core $R_{c}^{th}$, the valence neutron $R_{n}^{th}$ and total
      matter $R_{m}^{th}$ in
      $^{22}$N.  $S_{n}$ denotes the corresponding
      one-neutron separation energy.}
  \begin{tabular}{l    c  c  c  c    }
    \hline
    \hline
     &$S_{n}$   &$R_{c}^{th}$ &$R_{n}^{th}$ &$R_{m}^{th}$  \\
         &(MeV)       &(fm) &(fm) &(fm) \\
   \hline
  Set A   &1.490 &3.21 &5.25 &3.53  \\
  Set B   &1.280  &3.22 &5.47 &3.58  \\
  Set C   &1.070 &3.24 &5.75 &3.65  \\
    \hline
    \hline
  \end{tabular}
  \end{center}
\end{table*}

With the appropriate parameterization for the effective $^{21}$N-n interaction obtained
above, we calculate the bound-state observables of ${}^{23}$N in the $^{21}$N+n+n three-body
 model using the Faddeev formalism.  In these calculations, we use the cutoff 
$K_{max}$=20 in the hyperangular-momentum expansions
in Eq~\ref{eq:hyperspherical wave function} to provide a proper numerical accuracy. 
Comparing with the results of $K_{max}$=10, the difference between the two-body separation energies 
is about 4\%, which is much smaller than the relative error of about 30\% due
to different choices of parameters. Thus, this cutoff $K_{max}$=20 is adequate
 for our Faddeev calculations of the three-body system.
 We calculate the two-neutron separation energies $S_{2n}$
and r.m.s. matter radii of the ground and excited states of $^{23}$N using 
different sets of parameterization, and list the results in Table \ref{table:results}.  
We observe that both the the ground- and excited-state $S_{2n}$
decrease when the depth of the $^{21}$N-n potential increases.
We also find that the r.m.s. matter radii of the two states 
increase respectively with the decrease of $S_{2n}$ in 
the corresponding states.  This correlation which show 
that $^{23}$N is a standard halo nucleus,
as discussed in \cite{Ershov2012}. Our obtained two-neutron
separation energy of the $\frac{1}{2}^{-}$ ground state, $S_{2n}=3.6\pm0.5$ MeV, agrees 
well with the experimental result, {\it i.e.}, $S_{2n}=3.07\pm0.31$ MeV \cite{3Audi2012}. 
The r.m.s. matter radius of the ground-state $^{23}$N is
about 3.0 fm, which corresponds to the size of a stable nucleus with
mass number $A\approx27$. This radius is much smaller than $^{22}$C,
whose r.m.s. radius is about $5.4\pm0.9$ fm \cite{Tanaka2010}.
Therefore, although loosely-bound compared with stable nuclei, the halo structure of 
the $^{23}$N ground state may not be as clear as its isotone halo nucleus $^{22}$C.

\begin{table*}[htbp]
  \begin{center}
    \caption{\label{table:results}The two-neutron separation energies $S_{2n}$ and r.m.s. matter 
      radii $r_m$ of the ground and excited states of 
      $^{23}$N, calculated with three sets of parametrization and from the experiments. 
      The superscript $*$ denotes the excited state. }
  \begin{tabular}{l   c c  c  c  }
    \hline
    \hline
             &$S_{2n}$   &$r_m$   &$S_{2n}^{*}$  &$r_m^{*}$  \\
                       &(MeV)     &(fm)    &(MeV)         &(fm)  \\
   \hline
   Set A  &4.13  &2.969  &0.315 &4.272 \\
   Set B  &3.64  &2.985  &0.185 &4.358 \\
   Set C  &3.13  &3.004  &0.069 &4.476     \\
   Exp \cite{3Audi2012}    &3.07$\pm$0.31 &- &- &-        \\
    \hline
    \hline
  \end{tabular}
  \end{center}
\end{table*}

The $\frac{1}{2}^{-}$ excited state of $^{23}$N, 
which has not been discovered in experiments yet, 
is predicted in this work to have an extremely shallow
two-neutron separation energy $S_{2n}^{*}$. With different parameterization, we obtain
$S_{2n}^{*}$ to be in the region of 0.069--0.315 MeV, with a mean
value of 0.185 MeV.  The r.m.s. radii of this excited state
$r_m^{*}$ is about 4.3 fm, which corresponds to
the size of a stable nucleus with $A\approx$80.  Therefore, the
excited state of $^{23}$N can be unambiguously described as a giant halo state. 

To illustrate the structure of $^{23}$N, we calculate the
average distances between the two valence neutrons $r_{nn}$ and the
average distances from the core to the center-of-mass
of the valence-neutron pair $r_{C(nn)}$ in the $^{21}$N+n+n system. 
The results in Table \ref{table:radii} indicate that both ground and excited
states of $^{23}$N have triangular shapes.  By calculating 
the ratios of $r_{nn}$ and $r_{C(nn)}$ respectively in the ground and excited states, we find that
\begin{equation}
\label{eq:rnn-rCnn}
\frac{r_{nn}}{r_{C(nn)}} \approx \frac{r_{nn}^*}{r_{C(nn)}^*} \approx 1.9~.
\end{equation}
The same ratio obtained in both states suggests that these two
states have similar geometric structures and are mainly different by a spatially discrete scaling factor.

\begin{figure}[htbp]
  \centering
  \includegraphics[width=0.80\textwidth]{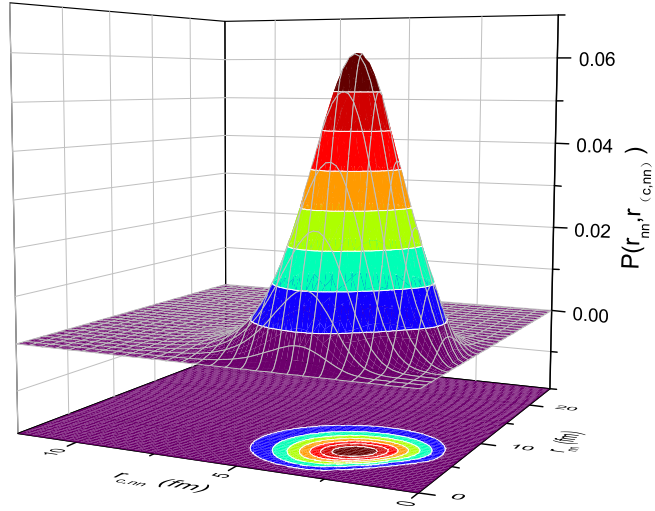}
  \caption{\label{fig:ground state} Contour diagram for the spatial
    correlation density distribution of the ground state of $^{23}$N
    with parameters of Set B. (Color online)}
\end{figure}

\begin{figure}[htbp]
  \centering
  \includegraphics[width=0.80\textwidth]{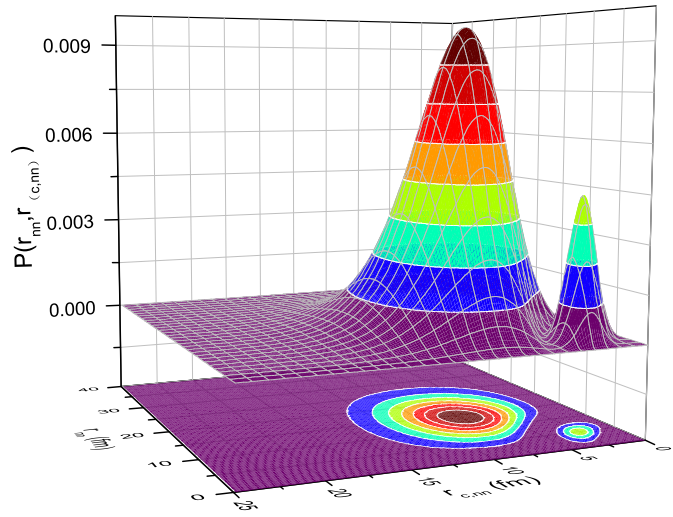}
  \caption{\label{fig:excited state} Contour diagram for the spatial
    correlation density distribution of the $\frac{1}{2}^-$ excited
    state of $^{23}$N with parameters of Set B. (Color online)}
\end{figure}

\begin{table*}[htbp]
  \begin{center}
    \caption{\label{table:radii} The
average distances between the two valence neutrons $r_{nn}$ and the
average distances from the core to the center-of-mass
of the valence-neutron pair $r_{C(nn)}$ in the $^{23}$N ground and excited states 
The superscript $*$ denotes the excited state.}
  \begin{tabular}{l    c  c  c   c }
    \hline
    \hline
         &$r_{nn}$ &$r_{C(nn)}$   &$r_{nn}^{*}$  &$r_{C(nn)}^{*}$ \\
         &(fm)     &(fm)                  &(fm)          &(fm)            \\
   \hline
   Set A   &6.415   &3.357   &14.600 &7.640   \\
   Set B   &6.681  &3.496    &16.326 &8.543   \\
   Set C   &6.878  &3.599    &17.059 &8.927       \\
    \hline  
    \hline
  \end{tabular}
  \end{center}
\end{table*}
Furthermore, we also analyze the correlation density distributions of the ground and 
excited states of $^{23}$N in configuration space.  These quantities are evaluated in
the Jacobi-coordinate representation with $^{21}$N as the spectator particle,
{\it i.e.}, in the third diagram in Fig.~\ref{fig:Jacobi
  coordinates}. The spatial correlation density distribution is
calculated as \cite{Romero2008}
\begin{equation}
\label{eq:Distribution} 
P(r_{nn},r_{C,(nn)})\equiv x^{2}y^{2}\int |\Psi_{i}^{JM}(x,y)|^{2}d\Omega_{x_{i}}d\Omega_{y_{i}}.
\end{equation}
The correlation densities for the ground and excited states are shown
in Figs.~\ref{fig:ground state} and \ref{fig:excited state}, respectively.
On the one hand, we find that
the ground state has the largest probability density when the two neutrons are
at a distance of about 3.5 fm from the core.  On the other hand, the excited state has
the largest probability density when the two neutrons are far away from the
core with a distance of about 8.5 fm, and are separated from each other
with a distance of about 16 fm.  The spatial separation in the excited state
of $^{23}$N supports the halo structure suggested above. 
Furthermore, in Fig.~\ref{fig:excited state},
a second peak with a smaller amplitude, which mainly comes from the $(0d)^{2}$
component, also appears in the excited state.
The occupancies of the valence neutrons are also calculated.  For
the ground state, the occupancies of the two valence neutrons in the
$(1s)^{2}$ and $(0d)^{2}$ states are 95\% and 5\%, respectively.
For the excited state, the occupancies of the two valence neutrons in the
$(1s)^{2}$ and $(0d)^{2}$ states are 77\% and 23\%, respectively.

Since the excited state has a much shallower two-neutron separation energy, a much larger r.m.s. radius, 
and much more extended spatial distribution than the ground state, one may be able to describe the excited 
state as an Efimov state in the halo nucleus~\cite{2Ren1994,Ren1996,Lin2012}. 
In fact, one can utilize the geometrical similarity between
the ground- and excited-state configurations to distinguish Efimov states~\cite{Nielsen1998}:
One of the magic properties of Efimov physics is
that two consecutive Efimov states can be related to each other by a discrete spatial scaling
factor~\cite{Nielsen1998}. In $^{23}$N, the dominant parts of the spatial density distribution
of the ground and excited states (Figs.~\ref{fig:ground state} and \ref{fig:excited state}) 
indicate that the  configurations of $^{23}$N in these two states have very similar shapes. 
Moreover, the same ratio between $r_{nn}$ and $r_{C(nn)}$ in the ground and excited states 
(Eq.~\eqref{eq:rnn-rCnn}) suggests that a discrete scaling symmetry exists between the two states.
These features are highly consistent with the explanation of Efimov states.
However, further theoretical or experimental investigations are needed to
confirm the excited state in $^{23}$N.

\section{Conclusion and discussions}
\label{sec:con_dis}
We have investigated the properties of $^{23}$N as a
three-body system consisting of an inert $^{21}$N core and two valence
neutrons. We apply the Woods-Saxon potential form to represent the effective 
$^{21}$N-n interaction, and reproduce the basic characteristics of the $^{22}$N.  
By solving the Faddeev equations of the $^{21}$N+n+n three-body system,
implemented with the $^{21}$N-n and n-n potentials, 
we calculate the bound-state observables in the $^{23}$N ground state, 
where we obtain the two-neutron separation energy $S_{2n}$ consistent
with the experimental result.
We also discover an excited $^{23}$N with a much shallower two-neutron separation energy of 0.18 MeV.
We also calculate the r.m.s. matter radii for the two states.  The obtained results suggest
a relatively small halo structure for the ground state. On the other hand, a more extended
distribution of the valance neutrons is unveiled in the excited state,
which indicates an distinct halo structure.  The average distances
between the valence neutrons, the average distances from the core
to the center-of-mass of the valence-neutron pair, and the spatial
correlation density distributions of the two states in $^{23}$N are also
calculated to investigate the geometric structure of the three-body
system. These results indicate that the two states have very similar
triangular shapes and can be related to each other by a discrete scaling symmetry. These features
suggest that the excited state of $^{23}$N can be an Efimov state.
Our calculations of $^{23}$N with the Faddeev formalism enable a new exploration
toward the neutron drip line, and urge further experimental investigation of the $^{23}$N 
ground state and the potential discovery of the $^{23}$N excited state.

\section*{Acknowledgment}
This work was supported by the National Natural Science Foundation of
China (grant nos 11035001, 11375086, 11105079, and 10975072), by the
973 National Major State Basic Research and Development of China
(grant nos 2013CB834400 and 2010CB327803), by the CAS Knowledge
Innovation Project no. KJCX2-SW-N02, by the Research Fund of Doctoral
Point (RFDP), grant no. 20100091110028, and by the Science and
Technology Development Fund of Macau under grant no. 068/2011/A. 
It was also supported in part by the Natural Sciences
and Engineering Research Council (NSERC) and the National Research
Council Canada.

\end{document}